\documentclass[twoside,twocolumn,9pt]{article}
\usepackage{extsizes}
\usepackage[super,sort&compress,comma]{natbib} 
\usepackage[version=3]{mhchem}
\usepackage[left=1.5cm, right=1.5cm, top=1.785cm, bottom=2.0cm]{geometry}
\usepackage{balance}
\usepackage{times,mathptmx}
\usepackage{sectsty}
\usepackage{graphicx} 
\usepackage{lastpage}
\usepackage[format=plain,justification=justified,singlelinecheck=false,font={stretch=1.125,small,sf},labelfont=bf,labelsep=space]{caption}
\usepackage{float}
\usepackage{fancyhdr}
\usepackage{fnpos}
\usepackage[english]{babel}
\addto{\captionsenglish}{%
  
}
\usepackage{array}
\usepackage{droidsans}
\usepackage{charter}
\usepackage[T1]{fontenc}
\usepackage[usenames,dvipsnames]{xcolor}
\usepackage{setspace}
\usepackage[compact]{titlesec}
\usepackage{hyperref}
\usepackage{placeins}

\definecolor{cream}{RGB}{222,217,201}

\begin{document}



\makeFNbottom
\makeatletter
\renewcommand\LARGE{\@setfontsize\LARGE{15pt}{17}}
\renewcommand\Large{\@setfontsize\Large{12pt}{14}}
\renewcommand\large{\@setfontsize\large{10pt}{12}}
\renewcommand\footnotesize{\@setfontsize\footnotesize{7pt}{10}}
\makeatother

\renewcommand{\thefootnote}{\fnsymbol{footnote}}
\renewcommand\footnoterule{\vspace*{1pt}%
\color{cream}\hrule width 3.5in height 0.4pt \color{black}\vspace*{5pt}} 
\setcounter{secnumdepth}{5}

\makeatletter 
\renewcommand\@biblabel[1]{#1}            
\renewcommand\@makefntext[1]%
{\noindent\makebox[0pt][r]{\@thefnmark\,}#1}
\makeatother 
\renewcommand{\figurename}{\small{Fig.}~}
\sectionfont{\sffamily\Large}
\subsectionfont{\normalsize}
\subsubsectionfont{\bf}
\setstretch{1.125} 
\setlength{\skip\footins}{0.8cm}
\setlength{\footnotesep}{0.25cm}
\setlength{\jot}{10pt}
\titlespacing*{\section}{0pt}{4pt}{4pt}
\titlespacing*{\subsection}{0pt}{15pt}{1pt}

\fancyfoot{}
\fancyfoot[LO,RE]{\vspace{-7.1pt}\includegraphics[height=9pt]{head_foot/LF}}
\fancyfoot[CO]{\vspace{-7.1pt}\hspace{13.2cm}\includegraphics{head_foot/RF}}
\fancyfoot[CE]{\vspace{-7.2pt}\hspace{-14.2cm}\includegraphics{head_foot/RF}}
\fancyfoot[RO]{\footnotesize{\sffamily{1--\pageref{LastPage} ~\textbar  \hspace{2pt}\thepage}}}
\fancyfoot[LE]{\footnotesize{\sffamily{\thepage~\textbar\hspace{3.45cm} 1--\pageref{LastPage}}}}
\fancyhead{}
\renewcommand{\headrulewidth}{0pt} 
\renewcommand{\footrulewidth}{0pt}
\setlength{\arrayrulewidth}{1pt}
\setlength{\columnsep}{6.5mm}
\setlength\bibsep{1pt}

\makeatletter 
\newlength{\figrulesep} 
\setlength{\figrulesep}{0.5\textfloatsep} 

\newcommand{\topfigrule}{\vspace*{-1pt}%
\noindent{\color{cream}\rule[-\figrulesep]{\columnwidth}{1.5pt}} }

\newcommand{\botfigrule}{\vspace*{-2pt}%
\noindent{\color{cream}\rule[\figrulesep]{\columnwidth}{1.5pt}} }

\newcommand{\dblfigrule}{\vspace*{-1pt}%
\noindent{\color{cream}\rule[-\figrulesep]{\textwidth}{1.5pt}} }

\makeatother

\twocolumn[
\begin{@twocolumnfalse}
\sffamily
\begin{center}
\begin{tabular}{p{15.cm} }
\begin{center}
\noindent\LARGE{\textbf{Stability limits of elemental 2D metals in graphene pores$^\dag$}} \\
\vspace{0.3cm}

 \noindent\large{Janne Nevalaita and Pekka Koskinen$^{\ast}$} \\
\end{center}
\vspace{0.3cm} \\
\noindent\normalsize{Two-dimensional (2D) materials can be used as stabilizing templates for exotic nanostructures, including pore-stabilized, free-standing patches of elemental metal monolayers. Although these patches represent metal clusters under extreme conditions and are thus bound for investigations, they are poorly understood as their energetic stability trends and the most promising elements remain unknown. Here, using density-functional theory simulations and liquid drop model to explore the properties of 45 elemental metal candidates, we identify metals that enable the largest and most stable patches. Simulations show that pores can stabilize patches up to $\sim 8$~nm$^2$ areas and that the most prominent candidate in a graphene template is Cu. The results, which are generalizable to templates also beyond graphene, provide encouragement for further, even more resolute experimental pursuit of 2D metals.} \\

\end{tabular}
\end{center}

 \end{@twocolumnfalse} \vspace{0.6cm}
]

\renewcommand*\rmdefault{bch}\normalfont\upshape
\rmfamily
\section*{}
\vspace{-1cm}


\footnotetext{\textit{Nanoscience Center, Department of Physics, University of Jyv\"askyl\"a, 40014 Jyv\"askyl\"a, Finland. E-mail: pekka.koskinen@iki.fi}}

\footnotetext{\dag~Electronic supplementary information (ESI) available. See DOI: 10.1039/c9nr08533e}




Mainstream two-dimensional (2D) materials research has long focused on layered van der Waals materials~\cite{miro14,lin16,novoselov19}. These materials have rigid covalent in-plane bonding and weak van der Waals out-of-plane bonding. They are interesting due to simple fabrication~\cite{novoselov05}, intrinsic properties~\cite{geim09,novoselov16}, possibility of materials design~\cite{geim13,Cao2018}, and a myriad of applications~\cite{deng16,fiori14,xu13}. However, yet a nascent practice is to use these 2D materials as templates to stabilize other, more exotic nanostructures.

\begin{figure}[t]
    \centering
    \includegraphics[width=\columnwidth]{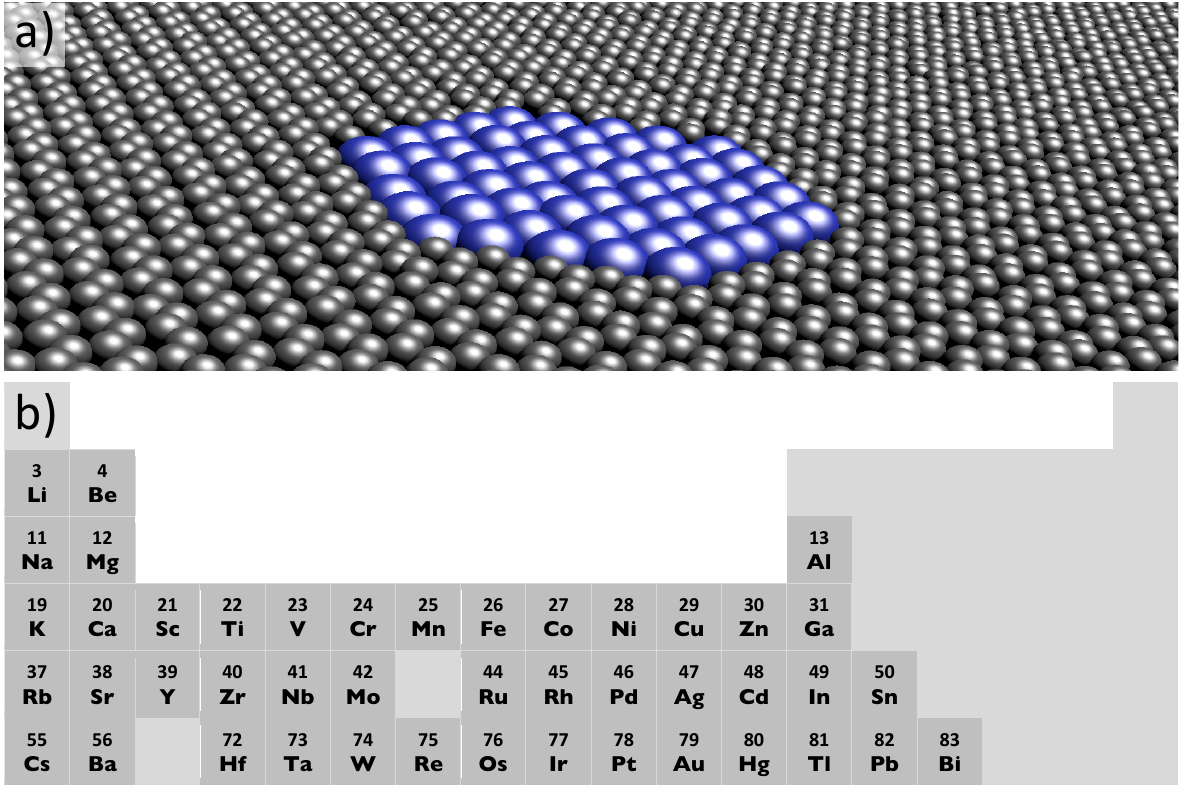}
    \caption{Elemental 2D metal patches in covalent templates. a) Schematic of an atomically thin, free-standing layer of close-packed elemental 2D metal monolayer patching a pore in graphene. b) The 45 elemental metal candidates explored in this work.}
    \label{fig:fig1}
\end{figure}

Stabilization occurs particularly well in pores, which are ubiquitous in 2D materials~\cite{russo12}. Pores enable stabilizing 2D nanostructures that otherwise would be unstable. Notably, pores in graphene have been used to stabilize even free-standing patches of 2D metals (Fig. \ref{fig:fig1}a)\cite{Wang2012a,zhao14,Ling2015a}. Here 2D metal patches mean monolayer clusters of elemental metals that are characterized by metallic bonding, homogeneous electron density, and superatomic states~\cite{nevalaita18,lin_PRL_09,lin_PRB_10,Stiehler2013}. As elemental metals usually prefer 3D clustering~\cite{koskinen07}, free-standing 2D metal patches represent materials under conditions so extraordinary that they are bound for investigations, as their properties and applications could differ markedly from the ones found in supported metal monolayers~\cite{yin09,ling15,chen18, ma18}.

However, synthesis and experimental control of 2D metal patches is challenging and still immature. For example, the usual mechanical exfoliation of 2D metals is unviable~\cite{coleman11,nicolosi13,Tang2014a}. Only few experiments have demonstrated inklings of success, reported as small graphene-stabilized patches of Fe~\cite{zhao14}, MoS$_2$-stabilized patches of Mo~\cite{zhao18}, as well as somewhat related graphene-stabilized patches of and Zn and Cu oxides~\cite{quang15, yin17}. Consequently, present research is best driven forward by simulations and modeling. Nevertheless, so far simulations have been limited to scattered elements and to effectively infinite membranes~\cite{yang15, yang15b, yang16, koskinen15, antikainen17,nevalaita18,nevalaita18b}. As a result, a practical and coherent understanding of the energetic trends in the patches as well as the most promising candidates among elemental metals remain unknown.

\begin{figure}[t]
    \centering
    \includegraphics[width=\columnwidth]{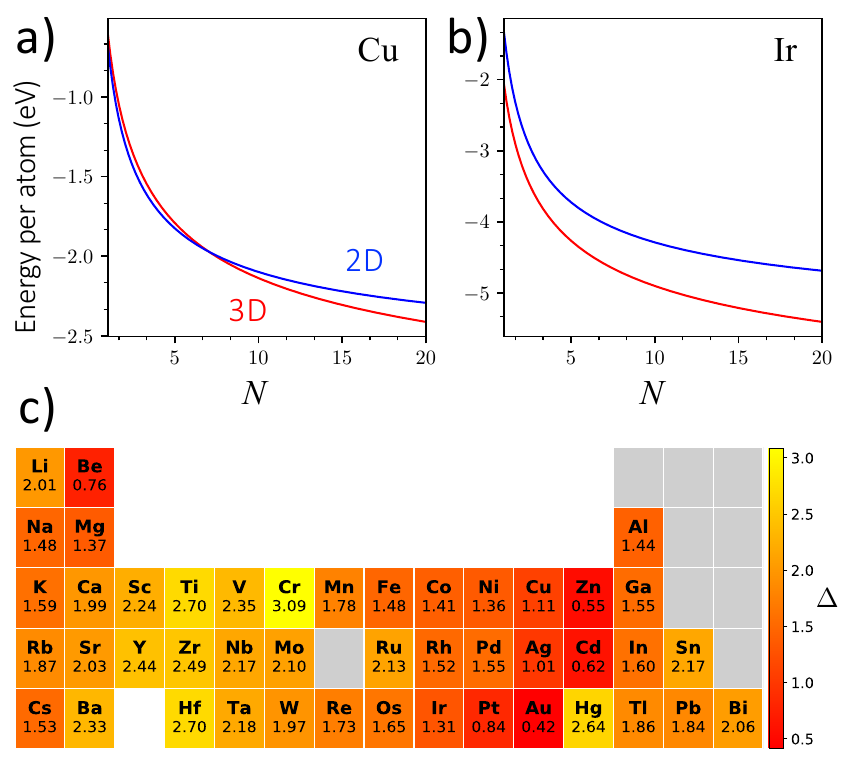}
\caption{Relative stability of 2D and 3D clusters in gas-phase, without the presence of stabilizing pores.  a) Energy per atom for 2D and 3D clusters of Cu according to the liquid drop model [eqs~(\ref{eq:E3D}) and (\ref{eq:E2D})]. b) Same as panel a for Ir. c) Dimensionless parameter $\Delta$ [eqn~(\ref{eq:delta})] visualized across the periodic table; smaller value stands for greater intrinsic 2D stability.}
    \label{fig:fig2}
\end{figure}

Therefore, in this work we address the fundamental question: \emph{Which elemental metals enable the most stable 2D patches in the pores of covalent 2D materials?} Using density-functional theory (DFT), we investigate finite patches of 2D metals and their interaction with a graphene template. The trends in key properties are explored across the periodic table, among 45 elemental metals (Fig.~\ref{fig:fig1}b). By combining DFT calculations with a liquid drop model, we are able to identify parts of the periodic table that hold the most prominent 2D metals. It turns out that, for graphene pores, the largest and most stable 2D metal patches are provided by Zn, Ag, and Au, the best candidate being Cu.

To investigate the energetic stability trends of 2D metal patches, we resort to the reliable liquid drop model~\cite{perdew91}. We begin by analyzing the stabilities of 2D and 3D clusters first in gas-phase, without the template. According to the model, the formation energy of a 3D cluster with $N$ atoms is
\begin{equation}
E(N) = \alpha V(N)+\sigma A(N) + E_c(N),
\end{equation}
where $V$ is the volume, $\alpha$ is the bulk energy density, $A$ is the surface area, $\sigma$ is the surface energy density, and $E_c$ is energy due to surface curvature. For simplicity, we choose closely packed, spherical clusters with radius $R$ and neglect the curvature energy term. Thus, the formation energy per atom for an $N$-atom cluster becomes
\begin{equation}
\varepsilon_\text{3D}^\text{gas}(N) = - \varepsilon^\text{3D}_\text{coh}+\sigma A(N)/N,
\label{eq:E3D}
\end{equation}
where $\varepsilon^\text{3D}_\text{coh}$ is the 3D bulk cohesion, $A(N)=c_\text{3D}d_\text{3D}^2 N^{2/3}$, $c_\text{3D}=(18\pi)^{1/3}\approx 3.84$, and $d_\text{3D}$ is the 3D bond length. Analogously, liquid drop model for the formation energy of a disc-shaped 2D patch is
\begin{equation}
\varepsilon_\text{2D}^\text{gas}(N) = - \varepsilon^\text{2D}_\text{coh}+\lambda L(N)/N,
\label{eq:E2D}
\end{equation}
where $\varepsilon^\text{2D}_\text{coh}$ is the 2D bulk cohesion, $\lambda$ is the edge energy, $L(N)=c_\text{2D}d_\text{2D}N^{1/2}$ is the length of the edge, $c_\text{2D}=(12 \pi^2)^{1/4}\approx 3.30$, and $d_\text{2D}$ is the 2D bond length. 

Since the 3D data for eqn~(\ref{eq:E3D}) can readily be found from the literature and the 2D data for eqn~(\ref{eq:E2D}) is found from our recent work \cite{nevalaita18,nevalaita18b}, we may proceed to investigate the relative stability of 2D and 3D gas-phase clusters quantitatively. Clearly, 3D clusters are more stable than 2D clusters and their relative stability increases faster upon increasing $N$ (Figs.~\ref{fig:fig2}a and \ref{fig:fig2}b). For some metals, such as Cu, crossover from 2D to 3D occurs at $N$ too small to have any quantitative meaning (Fig.~\ref{fig:fig2}a), while for most metals, such as Ir, crossover is absent altogether and the ground state is 3D for all $N$~(Fig.~\ref{fig:fig2}b). Not surprising, these trends are in accordance with the long known prevalence of 3D ground states in metal clusters~\cite{hakkinen02}.

However, despite the predictability of the 2D/3D energy ordering, it is illustrating to equate (\ref{eq:E3D}) and (\ref{eq:E2D}) to obtain an expression for the 2D/3D crossover. That expression
\begin{equation}
(\epsilon^\text{3D}_\text{coh}-\epsilon^\text{2D}_\text{coh})-\sigma c_\text{3D}d^2_\text{3D}N^{-1/3}+\lambda c_\text{2D} d_\text{2D}N^{-1/2}=0
\label{eq:cross}
\end{equation}
is simplified by two approximations. First, our previous work shows that for close-packed 2D metals $d_\text{3D}\approx d_\text{2D}$~\cite{nevalaita18}. Second, because close-packed structures are nearly isotropic, the surface energy and edge energy are related by $\sigma \approx \lambda/d_\text{2D}$. (See ESI\dag~for derivation of this relation.) Thus, the 2D/3D crossover occurs when
\begin{equation}
\Delta -c_\text{3D}N^{-1/3}+c_\text{2D}N^{-1/2}=0,
\label{eq:crossing}
\end{equation}
where 
\begin{equation}
\Delta = (\varepsilon^\text{3D}_\text{coh}-\varepsilon^\text{2D}_\text{coh})/(d_\text{2D}\lambda)
\label{eq:delta}
\end{equation}
is a dimensionless parameter, the smallness of which characterizes the intrinsic relative stability of 2D patches. 

The crossover occurs if Eq. (5) has a positive real solution, requiring $\Delta < 4/\sqrt{27}\approx 0.77$, which nominally holds only for few elements. Nevertheless, visualizing $\Delta$ across the periodic table gives us insight into elements' \emph{intrinsic} 2D stabilities (Fig. \ref{fig:fig2}c). As an intermediate result, most of the prominent candidates for elemental 2D metals are located at the end of the transition metal series. This trend is corroborated by the known disposition of Au towards 2D ground state geometries~\cite{hakkinen02,koskinen_NJP_06,koskinen07}.

The gas-phase results above provide a good starting reference, but our focus lies in the stabilization due to the interaction with pore edges. For concreteness, we will restrict ourselves to graphene pores~\cite{russo12}, because of graphene's mechanical rigidity and the feasibility of controlling pore sizes~\cite{Robertson2015,Wang2014a,kudin_PRB_01,koskinen_PRB_10b}.  

\begin{figure}[ht]
    \centering
    \includegraphics[width=0.9\columnwidth]{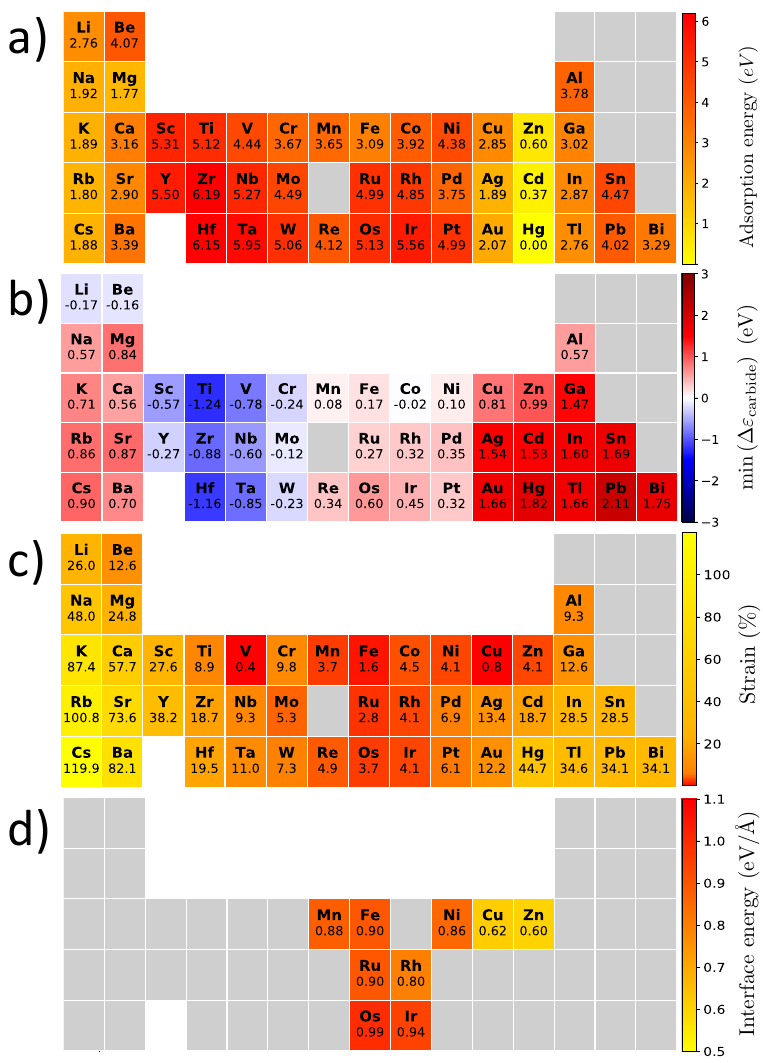}
    \caption{Viewpoints to interaction between elemental metals and graphene. a) Metal atom adsorption energies at graphene edges. The values shown are the minima among adsorption energies at hollow and bridge positions of armchair edges (Fig.~S3\dag). b) Carbide formation energies [eqn~(\ref{eq:ecarb})]. The values shown correspond to the minima of formation energies among three carbide configurations (Fig.~S5\dag). c) Absolute strains in 2D metals due to lattice mismatch relative to graphene. d) Interface energies between selected small-strain elemental 2D metals and graphene. The values shown are the minima among energies of four different metal-graphene interface configurations~(Fig.~S4\dag). }
    \label{fig:fig3}
\end{figure}

The stabilization of 2D clusters occurs only when the interaction between the metal atoms and the pore edge is exothermic and suitably directional. This requirement is fulfilled by graphene, which prefers in-plane adsorption at edges~\cite{Malola2009,antikainen17} as opposed to on-top adsorption~\cite{pastewka_JPS_13}. On-top adsorption energies, the values of which depend largely on the DFT functional used~\cite{pasti18}, remain below $2$~eV, mostly even below $0.2$~eV, whereas edge adsorption energies can be up to $7$~eV, depending on edge type and adsorption site (Figs.~\ref{fig:fig3}a and Fig.~S3; see ESI\dag~for computational details). These trends indicate that the nature of metal-graphene interaction at pore edges is well-suited for stabilizing 2D patches.

A caveat, however, is that the interaction and in-plane adsorption must not be too strong. Too strong and inapt interaction may cause graphene to swallow metal atoms altogether and give rise to metal carbides. We must therefore extend our investigations to various metal carbides and exclude elements that prefer carbide formation. To do this, we define carbide formation energy
\begin{equation}
    \Delta \varepsilon_\text{carbide}=[E_\text{carbide}-(N_C\mu_\text{gr}+N_M\mu_M)]/N_M,
    \label{eq:ecarb}
\end{equation}
where $E_\text{carbide}$ is the total energy of a carbide with $N_M$ metal atoms and $N_C$ C atoms, $\mu_\text{gr}$ is the chemical potential of C in graphene, and $\mu_M$ is the chemical potential of metal atoms in 2D bulk. $\Delta \varepsilon_\text{carbide}$ was calculated for all metals with three different carbide geometries (Fig.~S5\dag). Any occurrence of negative carbide formation energy implies that 2D metal-graphene interface is unstable; this consideration eliminates many of our candidates (Fig.~\ref{fig:fig3}b).

Above we considered edge adsorption of single atoms, which is different from the interaction between graphene and a finite 2D metal patch. To prevent interface buckling and the concomitant structural out-of-plane perturbations, interface should be commensurate whereby lattice mismatch between graphene and the metal patch should be minimal. A look at the matching of lattice constants thus provides yet another view into the most promising candidates for stable 2D patches with graphene (Fig.~\ref{fig:fig3}c). Incidentally, the picture of promising elements looks nearly the same for hexagonal boron nitride template, which has only $1.8$~\%\ lattice mismatch with graphene~\cite{Woods2014}.

Consequently, we chose nine metals with best-matching lattice constants for further analysis. Although V had the smallest lattice mismatch, it was dropped because of its negative carbide formation energy (Fig.~\ref{fig:fig3}b). We modeled the interface between metal and graphene using a ribbon model with different interface configurations (Fig.~S2\dag). Metal-graphene interface energy was defined as 
\begin{equation}
\lambda_\text{if} = (E_M+E_\text{gr}-E_\text{M+gr})/L_\text{if},
\end{equation}
where $E_M$ is the energy of the 2D metal ribbon, $E_\text{gr}$ is the energy of the graphene ribbon, $E_\text{M+gr}$ is the energy of fused ribbons, and $L_\text{if}$ is the length of the ribbon interface. Note that the interface energy entails the strain energy of the 2D metal. All of the resulting interface energies are positive, which indicates energetically stable metal-graphene interfaces (Fig.~\ref{fig:fig3}d); the dependence on the precise microscopic configuration was moderate (Fig.~S4\dag). Contrary to the intrinsic stability of gas-phase clusters (Fig.~\ref{fig:fig2}c), interface energies favor elements around the middle of the transition metal series.

\begin{figure}[ht]
    \centering
    \includegraphics[width=\columnwidth]{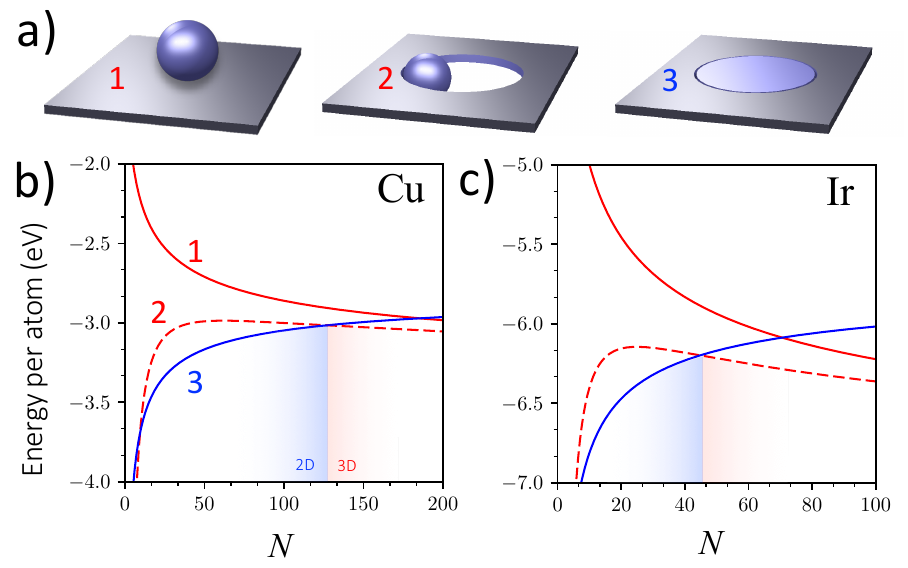}
    \caption{Relative stability of 2D and 3D clusters interacting with a graphene pore. a) Different interaction geometries: 3D clusters with on-top ($E_\text{ads}^{1}$) and pore edge ($E_\text{ads}^{2}$) adsorption, and 2D patches inside pores ($E_\text{ads}^{3}=\lambda_\text{if}L$). 
    b) Energy per atom for the three geometries for Cu ($E_\text{ads}^{1}=1$~eV, $E_\text{ads}^{2}=15$~eV, and $\lambda_\text{if}=0.62$~eV/\AA).
    c) Energy per atom for the three geometries for Ir ($E_\text{ads}^{1}=1$~eV, $E_\text{ads}^{2}=15$~eV, and $\lambda_\text{if}=0.94$~eV/\AA). The blue/red shading denotes the 2D/3D crossover.}
    \label{fig:fig4}
\end{figure}

We now have enough data to extend the liquid drop model to include interaction with graphene. This extension simply means the addition of cluster adsorption energies into eqs~(\ref{eq:E3D}) and (\ref{eq:E2D}), which yields the energy per atom
\begin{equation}
\varepsilon_\text{3D}(N) = - \varepsilon^\text{3D}_\text{coh}+\sigma A(N)/N + E_\text{ads}/N
\label{eq:E3D2}
\end{equation}
for 3D clusters and
\begin{equation}
\varepsilon_\text{2D}(N) = - \varepsilon^\text{2D}_\text{coh}+(\lambda-\lambda_\text{if})L(N)/N
\label{eq:E2D2}
\end{equation}
for 2D clusters, assuming that the edge and interface lengths are equal. Note that we assume a pre-existing pore and therefore omit energy contributions arising from the pore itself and that because the contribution from $E_\text{ads}$ scales as $\propto N^{-1}$, its influence at large $N$ is minor. Because $\lambda<\lambda_\text{if}$ for all metals (Fig.~\ref{fig:fig3}d), eqn~(\ref{eq:E2D2}) demonstrates explicitly the manner in which the pore stabilizes the patch with any finite $L$. When plotting the relative stability of 2D and 3D clusters using eqs~(\ref{eq:E3D2}) and (\ref{eq:E2D2}), the stabilizing effect of the pores becomes obvious (Figs.~\ref{fig:fig4}b and \ref{fig:fig4}c). For instance, 2D patches becomes energetically favored over 3D clusters with $N < 125$ for Cu and with $N < 45$ for Ir. The liquid drop model thus suggests that pores are able to stabilize 2D patches of respectable sizes.



\begin{figure}[!t]
    \centering
    \includegraphics[width=\columnwidth]{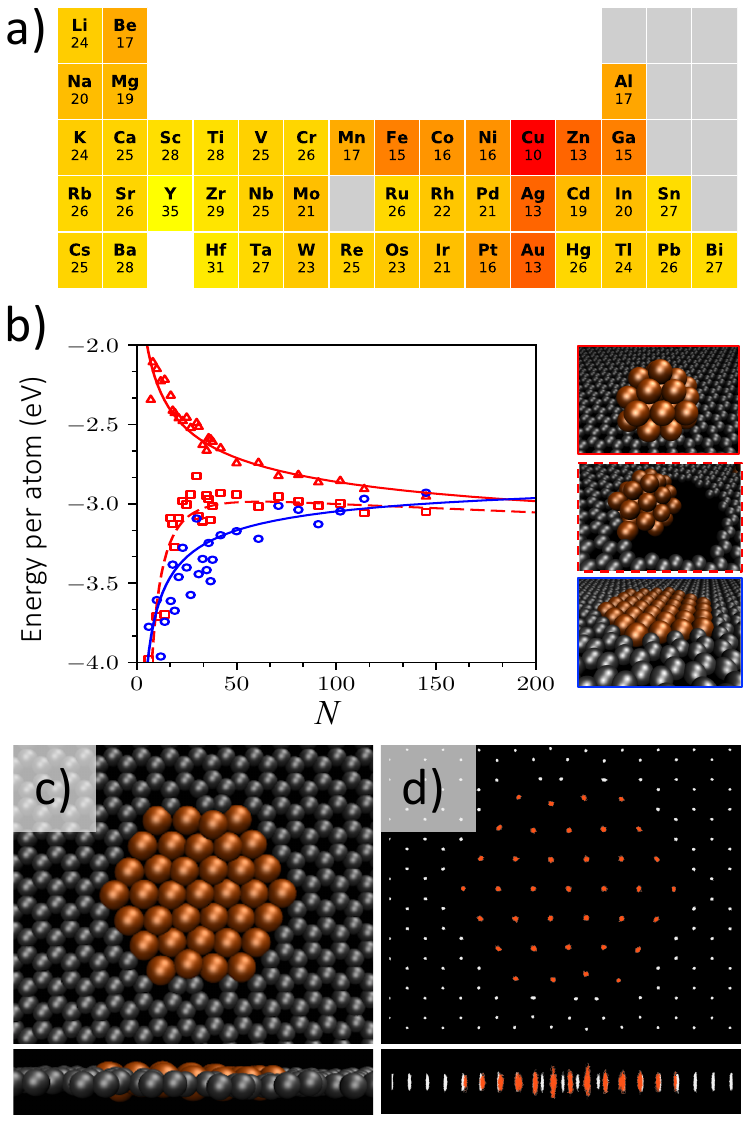}
    \caption{Identifying the best elements for stable 2D metal patches. a) Summative figures of merit of 2D stability for the 45 metal candidates. Smaller number means better 2D stability; numbers are rounded to closest integers.~(ESI\dag). b) Energy per atom for $N$-atom Cu clusters in three geometries: 3D clusters with on-top adsorption (red triangles; top panel on right), 3D clusters with pore edge adsorption (red squares; middle panel on right), and 2D patches with pore adsorption (blue circles; bottom panel on right). The lines represent the liquid drop model expressions (\ref{eq:E3D2}) and (\ref{eq:E2D2}) and symbols DFT values; the liquid drop model parameters are the same as in Fig.~\ref{fig:fig4}b. c) Snapshot views of a $37$-atom Cu patch in a graphene pore. d) Top and side views of Cu and C atom trajectories during a $36.9$~ps molecular dynamics simulation at $T=300$~K, demonstrating a kinetically stable patch. The snapshots in panel c are taken from this trajectory.}
    \label{fig:fig5}
\end{figure}

To answer the the fundamental question about the best candidates for stable 2D metal patches, let us then summarize our findings. We have investigated stability of 2D patches from four viewpoints: \emph{i}) intrinsic relative stability of the 2D metal (Fig.~\ref{fig:fig2}c), \emph{ii}) commensurability and the resulting strain at the interface (Fig.~\ref{fig:fig3}c), \emph{iii}) possibility of carbide formation (Fig.~\ref{fig:fig3}b), and \emph{iv}) the chemical bonding at the interface (Fig.~\ref{fig:fig3}a and \ref{fig:fig3}d). Because some viewpoints go beyond pure energetics, a balanced ranking of the elements must consider them simultaneously. We therefore united all the viewpoints into a single summative figure of merit, and ranked the metals accordingly (see ESI\dag~for description). According to the figure of merit, the foremost metals for stable patches are Zn, Ag, and Au, but the best one is Cu (Fig.~\ref{fig:fig5}a).


Adopting Cu as the best candidate, let us first go backwards to make a necessary validation of the extended liquid drop model. We created $27$ Cu clusters with $N=5\ldots 145$ and calculated the formation energies explicitly by DFT (ESI\dag). While the clusters do not
necessarily represent global energy minima, the energy differences between low-energy isomers are small enough (< 15 meV/atom) to have any influence on the energetic trends~\cite{grigoryan06,assadollahzadeh08,zhang19}. As before (Fig.~\ref{fig:fig4}a), we considered three different geometries: $1$) 3D clusters with on-top adsorption, $2$) 3D clusters with edge adsorption, and $3$) 2D patches in graphene pores (ESI\dag). The extended liquid drop model seems to capture the trends in the energetics reasonably well, even upon assuming an $N$-independent 3D adsorption energy~(Fig.~\ref{fig:fig5}b). Most important, DFT predicts energetically stable 2D patches up to $N\approx 120$, corresponding to an area of $\sim 8$~nm$^2$, and supports the predictions of the liquid drop model. While there are variations in the DFT energies due to differences in atomic arrangements, the fair agreement in Fig.~\ref{fig:fig5}b suggests that the trends are correct and indicates that stability could be achievable for respectable patch sizes; besides, the stability would even improve upon making the patches oblong.

In addition to pure energetics, we made preliminary investigation of kinetic stability. We conducted molecular dynamics simulation of an $N=37$ Cu patch for $36.9$~ps using a Langevin thermostat at $T=300$~K. In addition to corroborating previous simulations that have demonstrated kinetically stable infinite 2D Cu membrane~\cite{yang16}, our simulations suggest kinetically stable graphene-metal interface as well (see Figs.~\ref{fig:fig5}c and \ref{fig:fig5}d as well as Supplemental Movie ESI\dag).


To conclude, we have used liquid drop model in conjunction with DFT to show that pores in covalent 2D templates such as graphene could be used to stabilize 2D metal patches in respectable sizes. 
Although the best candidate, Cu, is familiar from graphene synthesis~\cite{Li2009b}, its prominence in this context is less clear. Anyhow, given the demonstrated usefulness of the liquid drop model, the results are swiftly generalizable to other covalent templates. The availability of free-standing, finite metal patches could serve as a platform for a number of applications including catalysis, sensing, bioimaging, photothermal therapy, solar cells, and electrical contacting~\cite{chen18}. Although we have predicted that the eventual patch structures ought to be stable, addressing the actual, experimental route to ultimately reach them is beyond our scope. However, pursuing this route resolutely is just what we legitimately may now propose.


\section*{Conflicts of interest}
There are no conflicts to declare.

\section*{Acknowledgements}
We acknowledge the Academy of Finland for funding (Project 297115).



\balance


\bibliographystyle{rsc}
\bibliography{apssamp} 

\end{document}